\documentclass[final,5p,times]{elsarticle}

\usepackage{amsmath}
\usepackage{amssymb}
\usepackage{booktabs}
\usepackage{capt-of}
\usepackage{enumitem}
\usepackage{graphicx}
\usepackage{microtype}
\usepackage{multirow}
\usepackage{placeins}
\usepackage{tabularx}
\usepackage{url}
\usepackage{xspace}
\usepackage{xcolor}
\usepackage{tikz}
\usepackage{siunitx}

\usetikzlibrary{arrows.meta, calc,fit,positioning}
\usepackage{booktabs,tabularx,float,array}

\journal{Future Generation Computer Systems}

\definecolor{cesblue}{RGB}{32,84,147}
\definecolor{cesgreen}{RGB}{46,116,78}
\definecolor{cesorange}{RGB}{180,82,27}

\newcommand{\RQ}[1]{\textbf{RQ#1}}
\newcolumntype{Y}{>{\raggedright\arraybackslash}X}
\newcolumntype{C}[1]{>{\centering\arraybackslash}p{#1}}

\setlist{nosep,leftmargin=*}
\begin{document}

\begin{frontmatter}

\title{CES: A Validated, Reproducible, and Scalable Simulator for the Edge-Cloud Continuum}

\author[oulu]{Nam Do\corref{cor1}}

\ead{Nam.Do@oulu.fi}
\author[oulu]{Rezwan Ahmad Nayreed}
\ead{RezwanAhmad.Nayreed@oulu.fi}
\author[oulu]{Henna Kokkonen}
\ead{henna.kokkonen@almpartners.fi}
\author[oulu]{Hassan Mehmood}
\ead{Hassan.Mehmood@oulu.fi}

\author[oulu]{Vu Truong}
\ead{Vu.Truong@oulu.fi}
\author[oulu]{Lauri Lov\'en}
\ead{Lauri.Loven@oulu.fi}

\cortext[cor1]{Corresponding author.}

\address[oulu]{Future Computing Group, University of Oulu, Finland}
\begin{abstract}
Evaluating edge-cloud systems on real hardware across many workloads and network conditions is expensive and hard to repeat. Discrete-event simulators offer a practical alternative, yet existing continuum simulators are rarely validated against physical systems, seldom verify reproducibility across different computing environments, and often slow down sharply at large scale. We present the Continuum Ecosystem Simulator (CES), a validated, reproducible, and scalable simulation platform for the edge-cloud continuum. CES combines a calibrated multi-worker server model, a network and HTTP model that accounts for uneven bandwidth sharing between overlapping transfers, and a managed workflow for running experiments locally or on Docker, Kubernetes, and HPC systems. In server experiments on a bare-metal HPC system across multiple workloads and worker counts, CES accurately predicts queuing and total server times on over 90\% of test runs, whereas PureEdgeSim and EdgeCloudSim miss queuing delays when calibrated only with task execution times. In end-to-end experiments on a cloud testbed, CES matches measured round-trip latencies and captures how simultaneous transfers finish spread out over time rather than all at once. Across local, containerized, cluster, and HPC environments, identical configurations produce byte-identical output files. Finally, at 30,000 simulated devices, CES runs nearly $20\times$ faster than PureEdgeSim and over $370\times$ faster than EdgeCloudSim. Source code and evaluation artifacts are openly available.
\end{abstract}


\end{frontmatter}

\section{Introduction}
\label{sec:introduction}
Modern applications increasingly distribute computation across IoT and mobile devices, edge servers, and remote cloud data centers, forming a unified computing continuum~\cite{shi2016edge,mach2017mobile,mao2017survey,kumar2022survey}. Deciding where each task executes directly shapes response time, network bandwidth usage, energy consumption, and monetary cost. Because continuum infrastructures combine heterogeneous hardware, time-varying workloads, and shared network links, evaluating placement and resource-management strategies systematically on physical testbeds is expensive, slow, and difficult to repeat. Discrete-event simulation therefore serves as a primary tool for conducting controlled continuum experiments at scale.
Despite wide adoption, existing edge-cloud continuum simulators face three practical limitations that restrict their credibility and usability. First, model fidelity is rarely validated against physical systems. Widely used simulators often rely on idealized execution and equal-bandwidth abstractions, omitting real-world overheads such as task dispatching, worker coordination, execution variability, and uneven concurrent network sharing that strongly influence queuing and tail latency. Second, execution reproducibility across environments is rarely verified. Moving experiments from a local workstation to containers, Kubernetes clusters, or HPC systems often makes results difficult to repeat unless configurations, software runtimes, and output artifacts are explicitly managed and tracked. Third, computational scalability degrades at large scale. As topologies grow to tens of thousands of devices and millions of tasks, topology setup and event-management overheads can slow simulations to minutes or hours, making large parameter sweeps impractical.
To address these gaps, we present the Continuum Ecosystem Simulator (CES), a validated, reproducible, and scalable discrete-event simulation platform for the edge-cloud continuum. Building on the software lineage of PureEdgeSim and EISim, CES combines calibrated server and network models with a managed, provenance-tracked experiment platform and a scalable simulation engine. Because claims of accuracy and reproducibility are meaningful only for specific workloads, hardware, and metrics, this paper evaluates CES empirically through three research questions:

\begin{description}[style=nextline]
\item[\RQ{1} -- Model fidelity.]
After calibration, how closely do CES predictions match new physical measurements of server execution and queuing times, and end-to-end latency under controlled network conditions? How closely do PureEdgeSim and EdgeCloudSim match the same server measurements when given the same calibrated inputs?
\item[\RQ{2} -- Execution reproducibility.]
Does the same versioned CES configuration produce identical outputs natively on a macOS workstation, in a Docker container, on a Kubernetes cluster, and on an HPC system, across different operating systems and processor architectures?
\item[\RQ{3} -- Computational performance and scalability.]
How quickly does CES produce complete simulation results compared with PureEdgeSim, EISim, and EdgeCloudSim under matched scenarios, and how does total wall-clock time scale as the number of devices and tasks increases? 
\end{description}
Relative to existing continuum simulators such as PureEdgeSim and EISim, this paper makes four contributions:
\begin{enumerate}
    \item \textbf{Calibrated multi-worker server lifecycle model (\RQ{1}):} A server model that tracks each task across a FIFO queue, a serial collector, and parallel workers, accounting for overlapping dispatch and execution as well as run-to-run execution variability. We validate it against held-out bare-metal HPC measurements across four workload types and multiple worker counts. 
    \item \textbf{Weighted concurrent network and HTTP end-to-end latency model (\RQ{1}):} A directed-link and HTTP transport model incorporating weighted bandwidth sharing for overlapping transfers, HTTP request slots, and response-write timing, validated against held-out end-to-end cloud measurements.
    \item \textbf{Managed, reproducible experiment platform (\RQ{2}):} An end-to-end submission, orchestration, monitoring, and artifact-archival workflow supporting GUI, API, and CLI clients across local, Docker, Kubernetes, and HPC environments, achieving byte-identical simulation outputs across operating systems and CPU architectures.
    \item \textbf{Scalable simulation engine (\RQ{3}):} Engine improvements including single-source route precomputation and constant-time entity registration that remove the setup bottlenecks of PureEdgeSim and EISim without changing simulation outputs. Avoiding both pairwise setup routing and the quadratic task-binding lookups seen in EdgeCloudSim keeps total runtime under one minute at 30,000 devices (nearly 20$\times$ faster than PureEdgeSim and over 370$\times$ faster than EdgeCloudSim).
\end{enumerate}
The remainder of the paper is organized as follows. Section~\ref{sec:related} reviews related continuum simulators, simulation validation, and reproducible experimentation. Section~\ref{sec:model} presents the CES server, network, and HTTP models, and Section~\ref{sec:architecture} describes the managed platform architecture. Section~\ref{sec:evaluation-methodology} details the experimental design and evaluation metrics for the three research questions, and Section~\ref{sec:evaluation-results} reports the results. Section~\ref{sec:discussion} discusses implications, limitations, and future work, and Section~\ref{sec:conclusion} concludes the paper.

\section{Related Work}
\label{sec:related}
\subsection{Continuum Simulation Platforms}
CloudSim provides a foundation for discrete-event simulation of cloud resources and resource-management policies~\cite{calheiros2011cloudsim}. iFogSim and EdgeCloudSim extend simulation towards distributed Edge and Fog environments, including models for deciding where applications run and how data moves between computing nodes, while EdgeCloudSim also supports device mobility \cite{gupta2017ifogsim,sonmez2018edgecloudsim}. iFogSim2 adds mobility, clustering, and microservice orchestration \cite{mahmud2022ifogsim2}. Recent comparative studies examine differences in the models, supported policies, and evaluation practices of Edge--Fog--Cloud simulators \cite{mechalikh2025quality,bendaouch2025benchmarking}. EdgeSimPy provides extensible Python components for studying Edge resource-management policies \cite{souza2023edgesimpy}. EdgeSim++ and Simcan2Fog provide detailed support for communication and network modeling \cite{lu2024edgesimpp,deaguilar2025simcan2fog}. The Computing Continuum Simulator connects infrastructure evaluation with deployment workflows, while ECLYPSE supports simulation and emulation within a common framework \cite{rodriguez2025continuum,massa2026eclypse}. MECSim extends continuum modeling with heterogeneous resources, dependent tasks, failures, and economic mechanisms
\cite{islam2026mecsim}. PureEdgeSim and EISim form the closest software lineage to CES \cite{mechalikh2021pureedgesim,kokkonen2023eisim}. PureEdgeSim supports heterogeneous Cloud, Edge, and Mist resources and evaluates resource management policies based on delay, energy consumption, resource utilization, and task outcomes. EISim extends this foundation with a deep reinforcement learning-based pricing method for dynamic resource pricing and price-aware task offloading across centralized, hybrid, and decentralized control topologies, as well as tools for agent training and evaluation. 
\subsection{Simulation Credibility}\label{sec:credibility}
Verification and validation address different aspects of a simulator's credibility: verification checks whether an implementation follows its declared model, whereas validation assesses whether that model adequately represents the target physical system~\cite{sargent2013verification}. While verification can be checked through unit tests, analytical checks, or internal consistency experiments, validation against physical hardware is substantially harder. It requires controlled physical testbeds and repeated trials to handle natural timing variation across runs. It must also capture real server and network behaviors that simplified models often leave out, such as task dispatch, worker coordination, multi-worker contention, and uneven TCP bandwidth sharing. 

As a result, most widely used fog, cloud, and edge simulation platforms focus on internal verification, trace-driven case studies, or relative policy comparisons, treating quantitative validation against physical testbeds as out of scope or future work. Specifically, CloudSim~\cite{calheiros2011cloudsim}, iFogSim~\cite{gupta2017ifogsim}, iFogSim2~\cite{mahmud2022ifogsim2}, EdgeCloudSim~\cite{sonmez2018edgecloudsim}, PureEdgeSim~\cite{mechalikh2021pureedgesim}, EISim~\cite{kokkonen2023eisim}, and EdgeSimPy~\cite{souza2023edgesimpy} are primarily evaluated in simulation rather than validated quantitatively against physical server and network measurements. For example, EdgeSimPy explicitly verifies selected model behaviors (data transfers, bandwidth sharing, and service migration) against its declared equations while leaving validation against physical systems as future work~\cite{souza2023edgesimpy}, and EISim similarly evaluates dynamic pricing and offloading policies in simulation without physical testbed validation~\cite{kokkonen2023eisim}. When unvalidated simulators are tested against real hardware, discrepancies can be substantial: Alshammari et al.\ reported RMS relative execution-time errors of 17.95\%-75.80\% when comparing CloudSim against distributed traffic and web-server workloads on Raspberry Pi~2 and~3 micro-datacenters~\cite{alshammari2017cloudsim}.

Only a smaller body of work directly validates discrete-event models against physical systems. SimGrid evaluates the accuracy and scalability of analytical computation and network-sharing models against real cluster measurements~\cite{casanova2014simgrid}, and Carvalho et al.\ reported close agreement between FogComputingSim simulations and physical measurements running EdgeBench~\cite{carvalho2022trust,das2018edgebench}. Together, these studies highlight that while model verification is common across continuum simulators, rigorous empirical validation against physical hardware remains scarce and is essential for credible simulation studies.

Calibration sets the values of a model's parameters for a particular target system. It is commonly understood as adjusting these parameters to maximize agreement between model outputs and observed data~\cite{trucano2006calibration}. This approach is useful when internal measurements are unavailable, instrumentation is costly, or the objective is to predict overall system performance under a specified set of operating conditions. A fitted model can still show strong predictive accuracy if it performs well on separate validation data. However, tuning parameters to match end-to-end outputs can make even an incomplete model fit the calibration data, so matching those data alone does not prove that the model captures the right underlying behavior. Parameters can instead be estimated from measurements of the system's individual components, with the simulator validated against separate end-to-end measurements. Dariol et al.\ use component measurements to calibrate analytical computation, communication, and power models with a SystemC framework. They evaluate the resulting predictions of 27 neural-network mappings, each tested with and without power management. This approach constrains component behavior before assessing whether the combined model reproduces system-level outcomes~\cite{dariol2026calibration}.  

Finally, credible validation requires quantitative agreement metrics over repeated independent trials rather than single-run comparisons, because real hardware exhibits multi-level timing variability~\cite{oberkampf2006measures,kalibera2013rigorous}. Hit-rate criteria combining relative and absolute tolerances provide a practical way to score repeated traces across wide latency ranges without over-penalizing small absolute errors~\cite{franke2008must}.

\subsection{Reproducible Continuum Experimentation}
Reproducible continuum experiments require the inputs, configuration, software environment, execution procedure, and analysis to be recorded and made available. The Popper convention applies version control and automated workflows to connect experiment execution, validation, analysis, and manuscript generation~\cite{jimenez2017popper}. Fogify supports repeatable emulation through containerized application descriptions, resource constraints, and scheduled network or fault scenarios. ExPECA combines automated resource provisioning with controlled wireless conditions to support repeatable physical measurements of edge-computing applications~\cite{symeonides2020fogify,mostafavi2023expeca}. E2Clab uses descriptions of infrastructure, services, networks, and workflow stages to automate deployment, execution, monitoring, and result collection. KheOps combines E2Clab with executable notebooks and artifact sharing to support the reuse of experiments across testbeds~\cite{rosendo2020e2clab,rosendo2023kheops}. More recently, Kiso has used declarative experiment descriptions to automate resource provisioning, software setup, execution, and output collection across heterogeneous testbeds~\cite{mayani2026kiso}. Recording provenance connects workflow inputs, processing steps, and results. ProvLight addresses the overhead of capturing this information on resource-constrained devices and integrates with E2Clab to support provenance capture across the edge-cloud continuum~\cite{rosendo2023provlight}. These approaches establish workflow automation and artifact preservation as important parts of continuum experimentation. 

\section{CES Platform Modeling}
\label{sec:model}
\subsection{Overview of Task Lifecycle}
\label{sec:task-lifecycle}
CES simulates tasks generated by edge devices that either run locally or are offloaded over a network to an edge or cloud server. For an offloaded task, the full lifecycle runs from when the client sends the request ($g_i$) until the client receives the complete response ($c_i$). The request first travels over the network to the server ($a_i$), where an HTTP layer accepts it and acquires one HTTP request slot before placing it into the server's FIFO queue (in standalone server tests without a network, tasks enter the FIFO queue directly).
Inside any edge or cloud server, three main components manage task execution: a \textbf{single FIFO queue}, a \textbf{serial collector}, and a \textbf{pool of single-threaded workers}. A task waits in the FIFO queue until both a worker slot and enough working memory are free. The serial collector then admits the task, reserves those resources, and sends the command to the assigned worker. The worker runs the application, prepares the result, and notifies the collector. The result then waits for the serial collector to receive and verify it. Only after verification does the server free the worker slot and memory ($u_i$) so another task can use them. Finally, the HTTP layer writes the response to the network ($v_i$), frees the HTTP request slot ($w_i^{\mathrm{resp}}$), and the response travels back to the client ($c_i$). Table~\ref{tab:rq1-notation} summarizes the timing notation across the full lifecycle.

\begin{table}[t]
\centering
\caption{Task lifecycle timing notation for task $i$ across client, network, and server (seconds).}
\label{tab:rq1-notation}
\footnotesize
\setlength{\tabcolsep}{3pt}
\begin{tabularx}{\columnwidth}{@{}l>{\raggedright\arraybackslash}X@{}}
\toprule
\textbf{Time} & \textbf{Event} \\
\midrule
$g_i$ & Client sends the task request (scheduled submission time). \\
$a_i$ & Complete request arrives at the server (enters FIFO queue directly in standalone server tests, or enters the HTTP layer first). \\
$b_i$ & Collector dequeues task and reserves a worker slot and memory. \\
$d_i$ & Collector starts sending the command to the worker. \\
$\rho_i$ & Worker receives and decodes the command ($R_i=\rho_i-d_i$). \\
$s_i$ & Application starts ($H_i=s_i-\rho_i$). \\
$z_i$ & Collector's send call returns ($D_i=z_i-d_i$; may occur before or after $s_i$). \\
$f_i$ & Application finishes ($X_i=f_i-s_i$). \\
$t_i^{\mathrm{send}}$ & Worker starts sending its result after post-processing ($P_i=t_i^{\mathrm{send}}-f_i$). \\
$\nu_i$ & Result becomes ready for collection ($N_i=\nu_i-t_i^{\mathrm{send}}$). \\
$\theta_i$ & Collector selects the result and enters setup. \\
$k_i$ & Collector starts receiving the result. \\
$h_i$ & Collector finishes receiving the result ($V_i=h_i-k_i$). \\
$u_i$ & Verification ends; worker slot and memory are freed ($\Phi_i=u_i-h_i$). \\
$v_i$ & HTTP layer starts writing response body to the network ($\tau_i^{\mathrm{send}}=v_i-u_i$). \\
$w_i^{\mathrm{resp}}$ & HTTP response write ends; HTTP request slot is freed ($\tau_i^{\mathrm{write}}=w_i^{\mathrm{resp}}-v_i$). \\
$c_i$ & Client finishes receiving the complete response body. \\
\midrule
\textbf{Duration} & \textbf{Meaning} \\
\midrule
$\omega_i$ & Planned wake-up delay for a parked collector. \\
$A^{\mathrm{adm}}$ & Admission-step duration before reservation. \\
$I_i$ & Worker-slot idle age before admission ($I_i=b_i-u_{j(i)}^{\mathrm{prev}}$). \\
$K_i$ & Dispatch setup or bookkeeping before sending the command. \\
$T_i$ & Collector turnaround after the last result in a batch. \\
$Q_i$ & Server queue delay, from arrival to application start ($Q_i=s_i-a_i$). \\
$S_i$ & Server sojourn time, from arrival to application finish ($S_i=f_i-a_i$). \\
$D_i^{\mathrm{req}}, D_i^{\mathrm{res}}$ & Request upload ($a_i-g_i$) and response download ($c_i-v_i$) durations. \\
$L_i^{\mathrm{E2E}}$ & End-to-end latency, from client send to response receipt ($c_i-g_i$). \\
$L_i^{\mathrm{comm}}$ & Communication residual ($L_i^{\mathrm{E2E}}-Q_i-X_i$). \\
\bottomrule
\end{tabularx}
\end{table}

\subsection{Server Task Lifecycle Modeling}
\label{sec:lifecycle}
\paragraph{Waiting: arrival and FIFO order}
The task arrives at the server at time $a_i$ and then is appended to the end of the waiting queue and enters \texttt{WAITING} state. No worker slot has yet been assigned to this task. When the serial collector has no work, it stays parked (idle) and can be woken up by either a new task arrival or a ready result from a worker. The first arrival during an idle period starts a wake-up duration $\omega_i \ge 0$, planned to finish at $a_i + \omega_i$. However, if a completed task result becomes ready at time $\nu_j$ before that wake-up finishes, the result wakes the collector immediately. The collector therefore becomes active at: 
\begin{equation}
t_i^{\mathrm{wake}}
=
\min\!\left(a_i+\omega_i,\nu_j\right),
\end{equation}
or simply at $a_i + \omega_i$ if no result becomes ready in the meantime. 
\paragraph{Admission: reservation of a worker slot and memory}
A waiting task can then be admitted, meaning taken out of the FIFO queue and given a worker slot and working memory, when both a worker slot and enough memory are available. Let $Q(t)$ be the FIFO queue at time $t$, $|Q(t)|$ be the number of waiting tasks, $O(t)$ be the number of occupied worker slots, and $W$ be the total number of worker slots, $M_{\mathrm{free}}(t)$ be the available working memory, and $m_{\operatorname{head}(Q(t))}$ be the memory required by the task at the front of the queue. Admission can begin when: 
\begin{equation}
|Q(t)| > 0,
\qquad
O(t) < W,
\qquad
m_{\operatorname{head}(Q(t))}
\leq
M_{\mathrm{free}}(t).
\label{eq:admission-conditions}
\end{equation}
Before reserving resources, the collector spends a nonnegative admission overhead $A^{\mathrm{adm}}$. Once this overhead elapses, the collector scans the free worker slots in ascending order and admits as many tasks from the front of the queue as the available worker slots and memory allow.  

For task $i$, let $b_i$ denote its admission time. At $b_i$, the task is removed from the waiting queue, assigned to a free worker slot $j(i)$, and moved to the $\texttt{RESERVED}$ state. From this point, the task holds the worker slot $j(i)$ and $m_i$ units of working memory. These resources are not freed immediately once the application finishes running. Instead, they stay reserved until the collector receives and verifies the result (checking its task ID, output size, and checksum for integrity), and are released only at time $u_i$. This rule matters for queuing because a task that has already finished running can still block the next waiting task from being admitted. 

After assigning the slot $j(i)$, the server records whether this is the first time the slot has been used. If the slot has been used before, it also records how long the slot sat idle since its previous task was released:
\begin{equation}
I_i
=
b_i-u_{j(i)}^{\mathrm{prev}},
\label{eq:worker-idle-age}
\end{equation}
where $u^{\mathrm{prev}}_{j(i)}$ is the resource-release time of the previous task on slot $j(i)$. Tracking first use and idle time matters in practice because a worker running its first task or waking up after sitting idle often behaves differently from a busy worker due to cold caches, runtime state, and CPU wake-up delays. CES later uses this history to choose the matching calibrated timings for dispatch, execution, and result preparation.

\paragraph{Dispatch: one serial collector and two concurrent timelines}
In the dispatch phase, the serial collector sends a command to each admitted task's assigned worker through its inter-process communication channel, providing the task ID and the information needed to start execution. Commands are sent one at a time, so the collector must finish its current send call before the next. Before each send begins, the model includes a measured setup or bookkeeping delay $K_i$ (measured from admission for the first task in a batch, or from the previous send return for later tasks in the same batch), during which the task is in the $\texttt{DISPATCHING}$ state. 

The collector begins sending task $i$'s command at time $d_i$, and the task enters $\texttt{HANDOFF}$. From this moment, the collector and the assigned worker advance on two independent timelines. On the collector timeline, the collector only waits for its \texttt{send} call to return (after duration $D_i$), not for the worker to finish the running task; once $\texttt{send}$ returns at $z_i = d_i + D_i$, the collector is free to send the next command, collect ready results, or admit new tasks. On the worker timeline, the worker does not have to wait for the collector's \texttt{send} call to return. Once it receives and decodes the command at $\rho_i = d_i + R_i$ and finishes any small pre-execution setup $H_i$, the application then starts at: 
\begin{equation}
    s_i = d_i + R_i + H_i
\end{equation}

Because the two timelines run independently, application start $s_i$ can happen either before or after the collector's send return $z_i$.

\paragraph{Application execution}
At time $s_i$, task $i$ enters the $\texttt{EXECUTING}$ state and the worker runs the task's actual application workload. If this execution takes duration $X_i$, the application finishes at $f_i$:

\begin{equation}
f_i=s_i+X_i
\end{equation}
CES models $X_i$ as a variable duration rather than a fixed constant because tasks running the same workload do not always have exactly the same execution time. The physical measurements contain variation at several levels, including \textbf{run-to-run variation, worker-level variation, task-specific variation}, and \textbf{worker-idle effects}. Even on the same workload on the same server, separate calibration runs can differ slightly in overall speed. Within the same run, different worker slots may have different average execution times. Even on the same worker, individual tasks can still finish faster or slower than the worker's average speed. For some workloads, the time for which a worker was idle before receiving the task may also be related to its execution time. 

Using one constant average execution time would remove those observed differences. It would also incorrectly treat every task and every worker as behaving identically. The model therefore represents the execution duration using an overall baseline together with a calibration trace effect, a worker effect, a task-specific effect, and an optional worker-idle effect. For task $i$, the execution duration is: 
\begin{equation}
    X_i=
\exp\!\left(
\mu+A_r+B_{r,j(i)}+\epsilon_i+G_i
\right)
\label{eq:ces-execution-time}
\end{equation}
A calibration trace is a recorded physical run of a workload on the target server. To capture the real run-to-run and worker-to-worker speed differences measured on hardware, CES selects one calibration trace $r$ for each simulated run. The term $\mu$ is the baseline  execution time on a logarithmic scale, calculated from the averages of the calibration traces of that workload and worker count. The trace effect $A_r$ captures how much trace $r$'s average execution time differs from $\mu$. The worker effect $B_{r,j(i)}$ captures how much worker slot $j(i)$ differs from the average of trace $r$. The residual $\epsilon_i$ captures the remaining task-specific variation after accounting for the trace and worker-slot effects. Finally, $G_i$ is an optional adjustment based on how long the assigned worker was idle before task $i$ was admitted. Because the terms inside parentheses are on a logarithmic scale, they represent the execution-time differences as relative changes rather than fixed additions in seconds. Taking the exponential converts the sum back into an execution duration in seconds. 
\paragraph{Worker post-processing and result readiness}
After the application finishes at time $f_i$, the worker does not send the result to the collector immediately. Instead, the task enters the $\texttt{WORKER\_POST}$ state while continuing to hold its worker slot and working memory. During this post-processing stage, the worker packages the result message and records runtime and integrity checks before replying (in our harness implementation for the physical server, this includes computing input and output hashes, checking CPU affinity, and recording changes in CPU, memory, scheduler, page-fault, context-switch, and garbage-collection counters). Letting $P_i$ denote this post-processing duration, the worker begins sending the result message at time $t_i^{\mathrm{send}}$:
\begin{equation}
t_i^{\mathrm{send}}=f_i+P_i    
\end{equation}

At $t_{i}^{\mathrm{send}}$, task $i$ leaves $\texttt{WORKER\_POST}$ and enters the $\texttt{RETURNING}$ state as the message travels over the inter-process communication channel. After a notification delay $N_i$ for the collector to detect the incoming result, the result becomes ready for collection at: 
\begin{equation}
\nu_i
=
t_i^{\mathrm{send}}+N_i
=
f_i+P_i+N_i.
\end{equation}

\paragraph{Result collection, verification, and release}
After task $i$'s result becomes ready at $\nu_i$, it waits in the result-ready queue until the collector checks for completed results. When the collector checks that queue, it takes all currently ready results as a single batch (any result that becomes ready later waits for the next batch) and processes them one at a time. When the collector starts handling task $i$ at time $\theta_i$, the task enters $\texttt{COLLECTOR\_SETUP}$ for a short setup or between-result handling step. At time $k_i$, the collector begins receiving and decoding task $i$'s result message, and the task enters the $\texttt{RECEIVING}$ state. Letting $V_i$ denote the receive duration, receipt finishes at: 
\begin{equation}
h_i=k_i+V_i
\end{equation}
After receipt finishes, task $i$ enters the $\texttt{VERIFYING}$ state. During this stage, the collector checks that the result belongs to the expected task and worker, matches the expected output size and checksum, and contains valid timing data. Letting $\Phi_i$ denote the verification duration, then resource release occurs at: 
\begin{equation}
u_i=h_i+\Phi_i
\end{equation}
The serial collector stays busy throughout result setup, receipt, and verification, finishing all three steps for the current result before moving to the next one. Meanwhile, other workers can continue running applications or preparing their results in parallel. At $u_i$, task $i$ enters $\texttt{RELEASED}$ state. The server frees worker slot $j(i)$, returns the reserved working memory, and records $u_i$ as the slot's most recent release time. If task $i$ is the last result in the selected batch, the collector also spends a turnaround time $T_i$ before it can start the next admission step, becoming available again at $u_i + T_i$. This completes the task's core execution lifecycle at the server. When tasks arrive over a network, the server's HTTP layer and the network model then deliver the response back to the client (Section~\ref{sec:transmission}, and~\ref{sec:transport}).

\begin{figure}[t]
\centering
\includegraphics[width=\columnwidth]{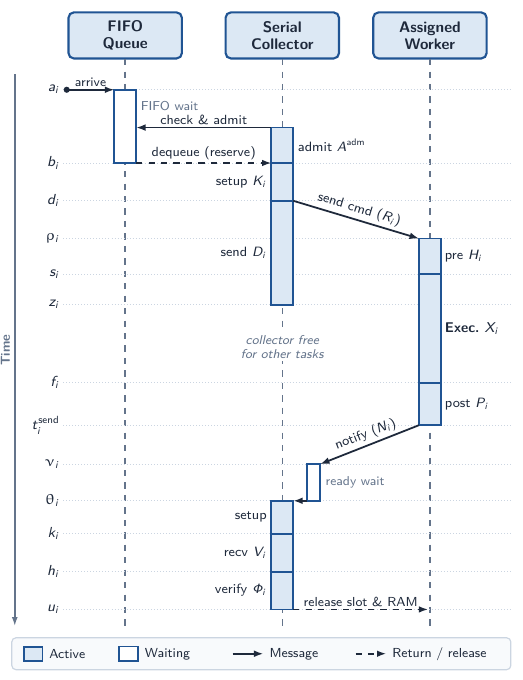}
\caption{Sequence diagram of one task's server lifecycle across the FIFO queue, serial collector, and assigned worker (notation in Table~\ref{tab:rq1-notation}). The worker slot and working memory are reserved upon dequeue at $b_i$ and released only after verification ends at $u_i$. Here the application starts ($s_i$) before the collector's send call returns ($z_i$). }
\label{fig:ces-server-task-sequence}
\end{figure}

\subsection{Network transmission and bandwidth sharing}
\label{sec:transmission}
Each directed link $\ell$ has an effective capacity $C_\ell$, measured in bits per second, and a configured delay $\delta_\ell$ measured in seconds. Upload and download use separate directed links and can have different capacities. Each route forwards complete transfers at each hop, meaning after transmission on a link finishes, its configured delay is applied before the transfer reaches the next node or link. 
\paragraph{Concurrent transfers} 
A flow is one request or response transfer currently being sent over a directed link. Let $\mathcal{A}_\ell(t)$ be the set of active flows on link $\ell$ at time $t$. CES allows multiple flows to overlap on the same link and supports both \textbf{equal bandwidth sharing} and \textbf{weighted bandwidth sharing}. Under weighted sharing, active flows do not have to split the link equally. When a transfer enters an idle link, CES draws a weight $w_v$ (using a deterministic seed) from a calibrated empirical set of weights and assigns it to that transfer. Any later transfer that arrives before the link becomes idle again receives weight $1$. Under equal sharing, every active transfer simply receives weight $1$. In both modes, the bandwidth allocated to active transfer $v$ on link $\ell$ is its share of total active weight, multiplied by the link capacity:  
\begin{equation}
r_{v\ell}
=
C_\ell\frac{w_v}{\sum_{z\in\mathcal{A}_\ell(t)} w_z}
\label{eq:canonical-flow-rate}
\end{equation}
where $C_\ell$ is the capacity of link $\ell$, $\mathcal{A}_\ell(t)$ is the set of active transfers on link $\ell$, $w_v$ is transfer $v$'s weight, and $r_{v\ell}$ is its allocated bandwidth.
\paragraph{Transmission progress} 
For each active transfer $v$ on link $\ell$, CES maintains a counter $B_{v\ell}$ of how many bits still remain to be transmitted and updates it at scheduled progress checks; the check interval is configurable. Over an elapsed interval $\Delta t$ at allocated rate $r_{v\ell}$, the remaining bits are updated as:  
\begin{equation}
B_{v\ell}^{\mathrm{new}}
=\max\!\left(0,
B_{v\ell}^{\mathrm{old}}-r_{v\ell}\Delta t\right).
\label{eq:canonical-network-progress}
\end{equation}

\subsection{HTTP transport to the server}
\label{sec:transport}
Let $a_i$ be the time when the complete request body reaches the server. Before submitting task $i$ to the server collector's FIFO queue, the server's HTTP layer must acquire one HTTP request slot (which caps how many requests the server handles at once; if all slots are occupied, incoming requests wait in FIFO order until a slot is freed). Once the collector finishes verifying task $i$ and releases its worker slot and memory at time $u_i$, the HTTP layer incurs a response-setup delay $\tau_i^{\mathrm{send}}$ before starting to write the response body to the client connection at time: 
\begin{equation}
v_i=u_i+\tau_i^{\mathrm{send}}.
\label{eq:canonical-http-send}
\end{equation}
At time $v_i$, the response enters the network model for delivery back to the client. In parallel on the server, the call that writes the response body to the client's HTTP socket takes duration $\tau_i^{\mathrm{write}}$ (drawn together with $\tau_i^{\mathrm{send}}$ as a calibrated pair) and finishes at:

\begin{equation}
w_i^{\mathrm{resp}}=v_i+\tau_i^{\mathrm{write}}
\label{eq:canonical-http-write}
\end{equation}

At $w_i^{\mathrm{resp}}$, the HTTP layer frees the HTTP request slot acquired earlier so another waiting request can enter. Because this local socket write can complete before the client receives the final response byte, $w_i^{\mathrm{resp}}$ governs server concurrency rather than adding directly to client-side download time.

\subsection{End-to-end latency}
The end-to-end latency is the duration from the scheduled submission time $g_i$ of the task request until the client has received the complete response body at $c_i$. We can define the request-path duration as $D_i^{\mathrm{req}} = a_i - g_i$ and response-path duration as $D_i^{\mathrm{res}} = c_i - v_i$. Then the complete latency decomposition can be expressed as: 

\begin{align}
L_i^{\mathrm{E2E}}
&= \underbrace{D_i^{\mathrm{req}}}_{\substack{\text{request}\\\text{upload}}}
 + \underbrace{(s_i - a_i)}_{\substack{\text{server}\\\text{queue delay}}}
 + \underbrace{X_i}_{\substack{\text{application}\\\text{execution}}} \nonumber\\
&\quad
 + \underbrace{(u_i - f_i)}_{\substack{\text{post-processing}\\\text{\& verification}}}
 + \underbrace{\tau_i^{\mathrm{send}}}_{\substack{\text{HTTP}\\\text{response setup}}}
 + \underbrace{D_i^{\mathrm{res}}}_{\substack{\text{response}\\\text{download}}}
\label{eq:canonical-e2e-decomposition}
\end{align}

This means the end-to-end latency timing components consist of request delivery, waiting until application execution starts, application execution, result preparation and collection through resource release, HTTP handling before response writing begins, and response body delivery to the client.

\section{CES Platform Architecture}
\label{sec:architecture}

\begin{figure}[H]
\centering
\includegraphics[width=\columnwidth]{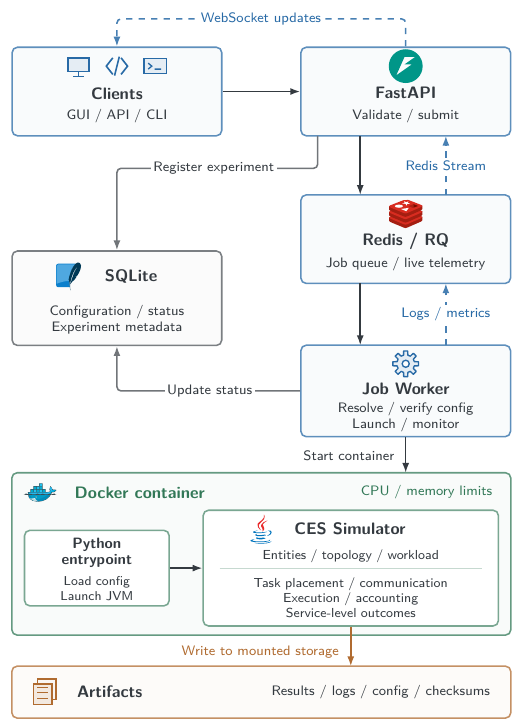}
\caption{CES platform architecture for the managed Docker execution path. Clients submit experiments through FastAPI, which registers metadata in SQLite and enqueues jobs in Redis/RQ. An RQ job worker verifies the resolved configuration, launches the containerized Python entrypoint and Java CES simulator with CPU/memory limits, streams live logs and metrics back via Redis Stream and WebSockets, and stores results and checksums on mounted storage.}
\label{fig:ces-platform-architecture}
\end{figure}

CES manages simulation experiments through a submission, execution, and result-retrieval workflow, as shown in Fig.~\ref{fig:ces-platform-architecture}. Users configure and submit experiments through a graphical interface or API clients, including command-line scripts. The FastAPI backend validates the request, assigns an experiment identifier, and records the configuration and initial status in SQLite. It then places the execution job in a Redis-backed RQ queue and returns the identifier for monitoring.

An RQ job worker retrieves the job and resolves the submitted settings into the complete simulation configuration. When a reviewed configuration hash is supplied, the job worker verifies that the resolved configuration matches it before execution. It saves the configuration, prepares output and model directories, and dispatches the experiment to the selected execution target. CES supports local execution, remote execution over SSH, and HPC execution through the Slurm workload manager with Apptainer containers. On the backend's own infrastructure, CES runs in a Docker container or in a container on a Kubernetes or OpenShift cluster. For the Docker path illustrated in the figure, the Docker runner creates a simulation container, applies CPU and memory limits, and mounts persistent input, output, and model directories. Remote execution collects outputs from the target system, while HPC execution additionally uses the scheduler to allocate resources and launch the simulation.

Inside the container, a Python entry point reads the configuration and launches the CES Java simulator. The simulator initializes the discrete-event engine and simulation manager, followed by the network model, continuum entities, pricing topology, task generator, and task orchestrator. It then processes scheduled events for task submission, network communication between entities, task placement, queuing, execution, and finally result delivery. Energy, carbon, economic accounting, and service-level outcomes are recorded during the simulation.

The Python RQ job worker monitors simulation output and publishes logs and metrics through Redis Stream. FastAPI forwards these updates to clients through WebSocket connections, while SQLite persists experiment status and metadata. Simulation results are written to mounted storage during execution. When the process exits, the job worker checks its exit code, updates the experiment status, and removes the container. For completed runs, it finalizes available evidence, including configuration and artifact checksums. The retained artifacts include result files, logs, the resolved configuration, and model checkpoints where applicable; users can inspect or download them through the backend.

\section{Evaluation Methodology}
\label{sec:evaluation-methodology}
Experiments on remote computing resources use services operated by CSC -- IT Center for Science in Finland: cPouta is an OpenStack-based cloud service on which we run virtual machines; Roihu is an HPC system whose bare-metal AMD EPYC 9965 compute nodes are allocated through Slurm and run containers with Apptainer; and Rahti is a Kubernetes-based container platform. Local experiments run on a macOS workstation (ARM64), both natively and in Docker. 
\subsection{RQ1: Model fidelity}

\subsubsection{Overview}
To answer RQ1, we test how closely CES task-timing predictions match measurements on real hardware. The evaluation has two parts. First, the \textbf{server-timing experiments} isolate server queuing, execution, and resource release across four workloads on bare-metal HPC nodes, with no network involved. Second, the \textbf{end-to-end experiments} add a controlled network and HTTP layer between two cloud VMs to evaluate full client-to-server-to-client latency under concurrent transfers. 

In both parts, we first run calibration traces (each trace is one experimental run under a fixed setting) to estimate model parameters, then compare CES predictions against measurements from separate held-out evaluation traces using predefined error metrics and pass criteria. For the server-timing experiments, we also run PureEdgeSim and EdgeCloudSim on the same calibration inputs, held-out traces, and pass criteria to compare CES against existing simulators. For end-to-end experiments, we compare CES's weighted bandwidth sharing against standard equal sharing. 

\subsubsection{Validation Hardware}

We use two physical systems for validation: Roihu compute nodes for the server-timing experiments and a cPouta client-server VM setup for the end-to-end experiments. Table~\ref{tab:rq1-resources} summarizes their
resource configurations.
\begin{table}[H]
\centering
\caption{RQ1 experimental systems and server resources.}
\label{tab:rq1-resources}
\small
\setlength{\tabcolsep}{3pt}
\renewcommand{\arraystretch}{1.1}
\begin{tabularx}{\linewidth}{
    @{}>{\raggedright\arraybackslash}p{0.26\linewidth}
    >{\raggedright\arraybackslash}X
    >{\raggedright\arraybackslash}X@{}}
\toprule
Resource & Roihu & cPouta \\
\midrule
Evaluation
& Server timing
& End-to-end latency \\

Platform
& HPC nodes
& Client/server VMs \\

Processor
& AMD EPYC, x86-64
& Intel Xeon (Cascadelake) \\

OS
& Linux
& Linux (guest OS) \\

Server CPUs
& 6 cores/job
& 4 or 6 vCPUs\textsuperscript{a} \\

Workers
& 1, 2, or 4
& 1, 2, or 4\textsuperscript{a} \\

Memory
& 8 GiB/job
& 7.57 or 15.25 GiB\textsuperscript{a} \\
\bottomrule
\end{tabularx}

\par\smallskip
\begin{minipage}{\linewidth}
\footnotesize
\textsuperscript{a}One- and two-worker configurations ran on a 4-vCPU server VM (7.57\,GiB guest-visible RAM); the four-worker object-retrieval configuration ran on a 6-vCPU \texttt{standard.extralarge} VM (15.25\,GiB guest-visible RAM).
\end{minipage}
\end{table}

\subsubsection{Workload Types and Parameters}
We use four different workloads: scalar arithmetic, image downsampling, data compression, and object retrieval with different input and output sizes. Each workload uses a deterministic input generator so that its input data and expected output can be reproduced. 

Scalar arithmetic processes 32 KiB of signed 32-bit integer readings, calculates their sum, count, minimum, and maximum. Image downsampling processes a $1024\times1024$ grayscale image and returns a $512\times512$ image. Data compression compresses a 1 MiB input using \texttt{zlib} at compression level 6. Object retrieval accepts a 64-byte request and generates a deterministic 1 MiB response in memory.

\subsubsection{Worker capacity and offered load}
Worker capacity $C_W$ is how many tasks per second that $W$ workers can complete together for a given workload and input size. Offered load $\rho$ tells us how quickly the experiment schedules incoming tasks relative to that capacity. If $\lambda$ is the scheduled average arrival rate in tasks per second, then:
\begin{equation}
\rho=\frac{\lambda}{C_W}, \qquad \lambda=\rho C_W
\end{equation}

\subsubsection{Server-Timing Calibration and Evaluation}
For each combination of workload, task specification, and worker count in Table~\ref{tab:server-calibration-profiles}, we build a calibration profile from physical measurements of worker and collector operations collected at offered load of $0.3$ (low arrival rate) and $1.1$ (high arrival rate). We then evaluate CES on separate held-out traces at an intermediate offered load of $0.8$ to test how accurately it predicts timing at a new arrival rate. Table~\ref{tab:rq1-calibration-evaluation} summarizes the trace and task counts for calibration and evaluation.

\begin{table}[t]
\centering
\caption{12 server timing calibration profiles. Each worker count
defines a separate profile for the stated workload.}
\label{tab:server-calibration-profiles}
\footnotesize
\setlength{\tabcolsep}{3pt}
\begin{tabularx}{\columnwidth}{@{}lXc@{}}
\toprule
\textbf{Workload} & \textbf{Task specification} & \textbf{Workers} \\
\midrule
Scalar arithmetic
  & 32\,KiB of signed 32-bit integers & 1, 2, 4 \\
Image downsampling
  & $1024 \times 1024$ grayscale pixels and an 8-byte header & 1, 2, 4 \\
Compression
  & 1\,MiB input, compressed with zlib at level~6 & 1, 2, 4 \\
Object retrieval
  & 64-byte request for a 1\,MiB object & 1, 2, 4 \\
\bottomrule
\end{tabularx}
\end{table}

During evaluation, CES loads the calibration profile matching the workload, task specification, and worker count, and inserts each task into the simulated server queue at the exact time $a_i$ when it entered the physical server queue. Rather than replacing each server operation with a single average, each profile stores empirical timing distributions built from the calibration traces, because real queuing and tail delays depend on timing variability as well as the mean speed. Since each simulation run samples one realization from these calibrated distributions (using a deterministic seed), we simulate each held-out trace 32 times with distinct fixed seeds to reduce sampling variance and obtain a stable estimate of the model's expected prediction. For each run, we compute the trace-level mean and nearest-rank 90th percentile (45th of 50 tasks) of three metrics, and average those statistics across all 32 runs to produce the predicted estimate $\hat{y}$:
\begin{itemize}
    \item \textbf{Queue delay $\hat{Q}_i$}: The time from queue arrival to application start, broken down using the lifecycle components from Section ~\ref{sec:lifecycle} as: 
    \begin{equation}
        \hat{Q}_i = (\hat{b}_i - a_i) + (\hat{d}_i -\hat{b}_i) + \hat{R}_i + \hat{H}_i = \hat{s}_i - a_i
    \end{equation}
    covering queue waiting until reservation, dispatch setup before sending the command, command delivery, and worker pre-execution setup.
    \item \textbf{Application execution duration $\hat{X}_i$}: Computed using the multi-level log-additive execution model in Eq.~\ref{eq:ces-execution-time}, capturing run-to-run, worker-slot, task-specific, and optional worker-idle variation.
    \item \textbf{Server sojourn time $\hat{S}_i$}: The total time from queue arrival to application finish, equal to the sum of predicted queue delay and execution duration:
    \begin{equation}
        \hat{S}_i = \hat{f}_i - a_i = \hat{Q}_i + \hat{X}_i
    \end{equation}
\end{itemize}

\begin{table}[t]
\centering
\caption{Calibration and held-out evaluation
design for each workload.}
\label{tab:rq1-calibration-evaluation}
\footnotesize
\setlength{\tabcolsep}{3pt}
\begin{tabular}{@{}lcccc@{}}
\toprule
Stage & Workers & \shortstack{Offered\\load}
& \shortstack{Traces per\\condition}
& \shortstack{Tasks per\\trace} \\
\midrule
Calibration & 1, 2 & 0.3, 1.1 & 10 & 200 \\
Calibration & 4    & 0.3, 1.1 & 10 & 500 \\
Evaluation  & 1, 2, 4 & 0.8  & 30 & 50 \\
\bottomrule
\end{tabular}

\smallskip
\begin{minipage}{\columnwidth}
\footnotesize
Each condition combines one worker count and one offered
load for a fixed workload. The design contains 60 ordinary
calibration traces and 90 evaluation traces per workload.
The evaluation traces contain 4,500 tasks per workload.
Dedicated component probes, instrumented controls, capacity
measurements, and diagnostic traces are additional and
are excluded from these counts.
\end{minipage}
\end{table}

\subsubsection{End-to-End Latency Calibration and Evaluation}
\label{sec:rq1-network-eval}
While the server-timing experiments isolate server-side timings, such as queuing or execution time, the end-to-end latency experiments evaluate whether CES predicts the full round-trip latency of a task across a controlled network path. The cloud testbed consists of two cPouta virtual machines (a client and a server) that exchange tasks over their private network. 

\paragraph{Controlled network path and server setup}
To keep the network bottleneck repeatable without modifying host network interfaces, both the client and server run inside isolated Linux network namespaces connected via virtual Ethernet interfaces (MTU 1500, with segmentation and receive offloads disabled). A Linux token-bucket filter (\texttt{tc tbf}) caps outgoing traffic on each namespace interface at $100\,\mathrm{Mbit/s}$, creating a symmetric $100\,\mathrm{Mbit/s}$ bottleneck in both directions. Traffic runs over the standard the Linux TCP CUBIC congestion control algorithm with $\texttt{TCP\_NODELAY}$ enabled, persistent HTTP/1.1 connections (120\,s keepalive).

We use a 4-vCPU server VM for the one- and two-worker configurations and a 6-vCPU server VM for the four-worker configuration. On the server, control threads (which manage HTTP connections, the serial collector, and timing logging) and worker processes (which run the task applications and record execution timings) are pinned to separate vCPUs so that control work does not compete with task execution or distort worker measurements. Specifically, one worker is pinned to CPU 0 and two workers to CPUs 0-1 (with control threads on CPUs 2-3), while four workers are pinned to CPU 0-3 (with control threads on CPUs 4-5). The server runs the same FIFO serial collector and persist worker pool as in the server-timing experiments, wrapped in an HTTP/1.1 server with eight HTTP request slots (limiting concurrent active requests to eight, with a 2 MiB maximum payload per request). Unlike the standalone server-timing experiments, full requests and response payloads are passed between the HTTP layer, collector, and workers, and both the collector and the client verify each response by byte length and SHA-256 checksum.

\paragraph{Burst traffic patterns}
To test different levels of network contention, both the bandwidth-sharing calibration and the held-out evaluation send 24-task traces organized into five burst patterns: (1) 1 request at a time, (2) 2 requests sent at the same time, (3) 2 requests sent $5\,\mathrm{ms}$ apart, (4) 4 requests sent at the same time, and (5) 4 requests sent $5\,\mathrm{ms}$ apart. Successive bursts within a trace are spaced far enough apart for each burst to finish before the next begins.

\paragraph{Calibration inputs}
To predict every term of Eq.~\eqref{eq:canonical-e2e-decomposition} without fitting parameters to end-to-end evaluation results, CES needs three pieces of information: how fast the link is for a single transfer, how overlapping transfers split the link, and how long a task spends inside the server. We measure these three inputs separately:
\begin{enumerate}
    \item \emph{Effective link capacity and baseline RTT:} Rather than assuming the nominal $100\,\mathrm{Mbit/s}$ shaping rate, we measure the usable application layer throughput separately in each direction using a dedicated network-probing endpoint (three repetitions of four concurrent 4\,MiB transfers per direction, with no application work). This yields effective rates of roughly $95.7\,\mathrm{Mbit/s}$ in each direction. We also measure host-to-host round-trip time (RTT) before and after each experiment run to confirm that network conditions remain stable. A fixed baseline round-trip delay of $0.612\,\mathrm{ms}$ (split evenly between upload and download links) accounts for path propagation and framing overhead.
    \item \emph{Bandwidth-sharing weights:} On a real TCP link, overlapping transfers rarely complete at the exact same moment. Typically the transfer that enters an idle link first receives a larger share of capacity. To model this behavior (Eq.~\eqref{eq:canonical-flow-rate}), we estimated an empirical set of 108 sharing weights once from a dedicated 15-trace (360-task) two-worker object-retrieval calibration run on the same $100\,\mathrm{Mbit/s}$ path, using 192-byte TCP header capture. Excluding 72 isolated single-flow transfers (which provide no information about bandwidth sharing), each of the 108 concurrent bursts was matched against eight candidate weights $\{1, 1.25, 1.5, 1.75, 2, 2.5, 3, 4\}$ by minimizing timing error across six intermediate transfer-progress checkpoints (10\%, 25\%, 50\%, 75\%, 90\%, and 100\% of response bytes). Rather than running calibration again for each setup, we reuse this same 108-weight distribution across all three worker counts as a transfer-level approximation of concurrent TCP sharing. 
    \item \emph{Per-configuration server and HTTP timing:} Because full-payload transfers and different VM sizes change internal server overhead, we collect separate server calibration traces on the target cPouta server VM for each workload and worker count. Each campaign starts with three short traces to measure baseline task execution time and set the calibration arrival rates, followed by 50 calibration traces, covering low and high open-loop arrival rates, repeated runs, and burst/idle tests (53 traces for one and two workers). For four workers on the 6-vCPU VM, we append six additional state-probe traces (59 traces total) and pool repeat execution timings per worker so all four slots have sufficient coverage. From these traces, we estimate the server lifecycle parameters (Section~\ref{sec:task-lifecycle}) together with the paired HTTP response-setup and socket-write durations $(\tau_i^{\mathrm{send}}, \tau_i^{\mathrm{write}})$. 
\end{enumerate}

\paragraph{Held-out evaluation}
To evaluate CES under concurrent network contention, we evaluate the object retrieval workload with 1, 2, and 4 workers, testing each on 25 held-out traces of 24 tasks each split evenly across the five burst patterns (Table~\ref{tab:rq1-network-design}). For each held-out trace, CES is run once with a fixed random seed, and all 25 predictions per worker count (five burst patterns $\times$ five repetitions) are generated and saved before running the physical traces. Physical held-out runs execute with packet capture turned off so that capture overhead cannot distort end-to-end latency. To isolate the effect of weighted sharing, we also compare against CES's equal-sharing mode on the same 25 held-out traces per worker count.

\begin{table}[t]
\centering
\caption{End-to-end network calibration and held-out evaluation design on cPouta.}
\label{tab:rq1-network-design}
\footnotesize
\setlength{\tabcolsep}{3pt}
\begin{tabularx}{\columnwidth}{@{}l>{\raggedright\arraybackslash}Xc@{}}
\toprule
\textbf{Stage} & \textbf{Protocol / patterns} & \textbf{Scale} \\
\midrule
Link-rate probes & 4 concurrent 4\,MiB transfers per direction (pre/post) & 3 repetitions/direction \\
Sharing weights & Two-worker retrieval bursts (192\,B TCP header capture) & 108 bursts \\
Server \& HTTP calibration & Baseline traces + arrival-rate, repeat, and burst/idle tests & 53--59 traces \\
Held-out evaluation & 5 burst patterns: $1$ request, $2$ requests (0/5\,ms), $4$ requests (0/5\,ms); no packet capture & 25 traces (600 tasks) \\
\bottomrule
\end{tabularx}
\end{table}

\subsubsection{Acceptance metric}

Repeated physical runs of the same configuration can yield different timings because of many factors, including system scheduling, worker wake-up, CPU frequency scaling, and so on. Moreover, a purely relative error bound could become overly sensitive when observed latencies are very small (tens of microseconds). We therefore evaluate agreement between CES predictions and physical measurements using an \textbf{absolute-or-relative acceptance} criterion of the hit-rate form introduced by Franke et al.~\cite{franke2008must}:
\begin{equation}
  |\hat{y} - y| \le \max\!\bigl(\tau_{\mathrm{abs}},\; \tau_{\mathrm{rel}}\,y\bigr),
  \label{eq:rq1-acceptance}
\end{equation}
where $y$ is an observed trace-level summary statistic, $\hat{y}$ is the corresponding CES prediction, $\tau_{\mathrm{rel}}$ is the relative tolerance, and $\tau_{\mathrm{abs}}$ is the absolute tolerance floor. 

For the \textbf{server-timing} experiments, we evaluate six trace-level summary statistics: the arithmetic mean and the nearest-rank 90th percentile of queue delay, execution duration, and application-finish sojourn time: 

\begin{equation}
  |\hat{y} - y| \le \max\!\bigl(200\,\mu\mathrm{s},\; 0.20\,y\bigr).
  \label{eq:rq1-server-acceptance}
\end{equation}
We chose the $200\,\mu\mathrm{s}$ floor based on earlier pilot sensitivity runs and fixed it before collecting held-out evaluation traces. To put physical run-to-run variability in context, we also ran a separate repeatability study of matched image-workload traces across eight independent Slurm allocations (2,304 ordinary 50-task traces), following the multi-level benchmarking principles of Kalibera and Jones~\cite{kalibera2013rigorous}. As shown in Table~\ref{tab:rq1-repeatability}, repeated runs with identical inputs and nominal schedules vary by tens to hundreds of microseconds at low load ($\rho = 0.3$) and by hundreds of microseconds to milliseconds under heavy queuing ($\rho = 1.1$), with larger variation across independent allocations than within a single allocation, which directly motivates us to use the absolute-relative tolerance metric.

\begin{table}[t]
\centering
\caption{Physical run-to-run repeatability standard deviation ($\mu$s) for matched image-workload traces across eight independent Slurm allocations on Roihu (2,304 ordinary 50-task traces; pair-based SD $= \sqrt{\frac{1}{2}\overline{\Delta^2}}$).}
\label{tab:rq1-repeatability}
\scriptsize
\setlength{\tabcolsep}{0pt}
\begin{tabular*}{\columnwidth}{@{\extracolsep{\fill}}ccrrrr@{}}
\toprule
& & \multicolumn{2}{c}{\textbf{Queue mean SD ($\mu$s)}}
& \multicolumn{2}{c}{\textbf{Execution mean SD ($\mu$s)}} \\
\cmidrule(lr){3-4}\cmidrule(l){5-6}
Load ($\rho$) & Workers & Within alloc. & Between alloc. & Within alloc. & Between alloc. \\
\midrule
0.3 & 1 & 89.3    & 151.1   & 75.7 & 110.8 \\
0.3 & 2 & 37.5    & 61.2    & 66.5 & 99.6  \\
0.3 & 4 & 24.1    & 33.0    & 62.1 & 96.8  \\
\midrule
1.1 & 1 & 1,849.7 & 3,249.9 & 77.5 & 100.8 \\
1.1 & 2 & 758.2   & 1,494.9 & 56.2 & 84.4  \\
1.1 & 4 & 407.5   & 886.4   & 52.6 & 90.3  \\
\bottomrule
\end{tabular*}
\end{table}

For the \textbf{end-to-end latency} experiments, we evaluate two per-task quantities: \textbf{end-to-end latency} $L_i^{\mathrm{E2E}} = c_i - g_i$ and the \textbf{communication residual }$L_i^{\mathrm{comm}} = L_i^{\mathrm{E2E}} - Q_i - X_i$, which isolates network transfer and HTTP/post-execution overhead from server queuing and application execution. On the physical testbed, $L_i^{\mathrm{E2E}}$ is measured strictly on the client clock while $Q_i$ and $X_i$ are measured strictly on the server monotonic clock; subtracting durations avoids any dependence on clock synchronization between VMs. Physical and simulated tasks are matched by task identifier and summarized per trace by their nearest-rank median (p50) and 90th percentile (p90), yielding four primary metrics. Applying Eq.~\eqref{eq:rq1-acceptance} to each trace, a metric passes when:
\begin{equation}
  |\hat{y} - y| \le
  \begin{cases}
    \max\!\bigl(30\,\mathrm{ms},\; 0.20\,y\bigr), & \text{for p50}, \\
    \max\!\bigl(75\,\mathrm{ms},\; 0.30\,y\bigr), & \text{for p90}.
  \end{cases}
  \label{eq:rq1-e2e-acceptance}
\end{equation}
The wider p90 bound reflects the higher sampling variability of tail quantiles on 24-task burst traces. In practice, the $30\,\mathrm{ms}$ (p50) and $75\,\mathrm{ms}$ (p90) floors apply when measured latencies are below $150\,\mathrm{ms}$ and $250\,\mathrm{ms}$, respectively; for longer latencies, the 20\% and 30\% relative limits take over.

\subsubsection{Comparison with PureEdgeSim and EdgeCloudSim}
To compare CES's server-timing fidelity against existing continuum simulators on identical physical measurements, we evaluate two baselines on the same held-out traces and workloads: PureEdgeSim (rev.~\texttt{bcdd239}), which shares the same architectural lineage as CES, and EdgeCloudSim (rev.~\texttt{83969dd}, bundled with CloudSim~7.0.0-alpha), which comes from a distinct CloudSim-based simulator family.
\paragraph{Native simulator configurations}
Each baseline runs its unchanged native scheduling engine with network, mobility, and placement layers bypassed so that tasks enter the server directly at their measured arrival time $a_i$. PureEdgeSim uses its default computing node with $W \in \{1, 2, 4\}$ cores (one core per task), native FIFO queuing, and nonbinding memory and storage. EdgeCloudSim uses its native \texttt{Task}, \texttt{EdgeVM}, and CloudSim \texttt{Datacenter}/\texttt{Broker} components configured with \texttt{CloudletSchedulerSpaceShared} ($W$ CPU cores, called processing elements in CloudSim, with one core per task) to match the non-preemptive physical worker pool. No CES collector or resource-release logic is added to either baseline.
\paragraph{Calibration inputs: execution-only and occupancy-calibrated baselines}
All baseline inputs are drawn exclusively from the same calibration traces ($\rho \in \{0.3, 1.1\}$) used for CES. As with CES, each held-out trace is simulated 32 times in both baselines using the same deterministic seeds, and the predicted trace statistics are averaged across all 32 runs. Because existing simulators release a core immediately upon task completion and do not separately model dispatch, handoff, post-processing, or collector verification, we evaluate two complementary baseline configurations:
\begin{itemize}
    \item \emph{Standard execution-only baseline:} Each task's native service demand encodes only its measured application execution duration $X_i = f_i - s_i$, using empirical calibration samples. Native start and finish directly define predicted application start $\hat{s}_i$ and finish $\hat{f}_i$. This configuration evaluates each simulator under standard out-of-the-box usage and isolates the effect of omitting post-execution slot occupancy.
    \item \emph{{Occupancy-calibrated baseline:}} To build a stronger baseline without modifying either simulator's scheduler, we sample paired triples ($B_i, \Delta_i^{\mathrm{start}}, X_i$) from the calibration traces ($\rho \in \{0.3, 1.1\}$), where $B_i = u_i - b_i$ is full worker-slot occupancy (from admission to resource release), $\Delta_i^{\mathrm{start}} = s_i - b_i$ is the admission-to-start offset, and $X_i = f_i - s_i$ is the application execution duration. Each task's native service demand encodes $B_i$, so the simulated core stays occupied until worker release at $\hat{b}_i + B_i$. From the simulated core-acquisition time $\hat{b}_i$, we then reconstruct predicted application start $\hat{s}_i = \hat{b}_i + \Delta_i^{\mathrm{start}}$ and finish $\hat{f}_i = \hat{s}_i + X_i$.
\end{itemize}
Because neither baseline accepts service durations in seconds directly, we convert each duration $t$ to a native task length using a fixed MIPS rate ($\mathrm{length} = \mathrm{MIPS} \times t$). PureEdgeSim accepts floating-point task lengths directly (at $10^6$~MIPS). Because EdgeCloudSim requires integer task lengths and inherits CloudSim's $0.01$-unit minimum completion-update step, we run it with a rescaled clock ($10^8$ simulation units per physical second, at $10^4$~MIPS) so that its minimum step corresponds to $0.1\,\mathrm{ns}$ rather than $10\,\mathrm{ms}$. Synthetic timing checks confirm that these conversions match the exact analytical FIFO schedule within $1\,\mu\mathrm{s}$.

\subsection{RQ2: Execution Reproducibility}
RQ2 tests whether the same simulation inputs produce identical outputs when CES runs in different execution environments. We run three configurations in four execution environments: natively on macOS (ARM64), in Docker (ARM64), on a CSC Rahti Kubernetes (x86-64) cluster, and on Roihu HPC (x86-64). Each configuration runs twice in each environment, giving 24 runs in total.

\subsubsection{Simulated Configuration}

All configurations contain 200 heterogeneous devices, 20 heterogeneous edge servers, and two cloud servers. Table~\ref{tab:rq2-entities} lists their processing and memory capacities. These are simulated capacities, separate from the physical resources used to run CES.

The experiments use centralized control, weighted task placement, fixed pricing, and transmission delays with equal sharing of link capacity. Each configuration uses a fixed task trace covering 300 simulated seconds. The entity capacities remain unchanged across all three configurations.
\begin{table}[t]
\centering
\caption{Simulated entities used in all RQ2 configurations.
Processing rates are per core; RAM values are in MB.}\label{tab:rq2-entities}
\small
\setlength{\tabcolsep}{3pt}
\renewcommand{\arraystretch}{1.15}
\begin{tabularx}{\linewidth}{
    @{}>{\raggedright\arraybackslash}X
    r r r r@{}}
\toprule
Entity class
& Count
& Cores
& \shortstack{MIPS\\per core}
& \shortstack{RAM\\(MB)} \\
\midrule
Lower-capacity smartphone
& 80 & 4 & 4,000 & 4,000 \\

Higher-capacity smartphone
& 60 & 6 & 6,000 & 6,000 \\

Tablet
& 40 & 4 & 3,000 & 2,000 \\

Laptop
& 20 & 6 & 7,000 & 8,000 \\
\midrule
Edge server A
& 6 & 20 & 23,809 & 115,663 \\

Edge server B
& 8 & 15 & 19,048 & 77,108 \\

Edge server C
& 6 & 10 & 14,286 & 48,193 \\
\midrule
Cloud server
& 2 & 64 & 50,000 & 262,144 \\
\bottomrule
\end{tabularx}
\end{table}

The three configurations vary network conditions and task load, as shown in Table~\ref{tab:rq2-configurations}. The temporal configurations replay bandwidth and delay measured with ping and iperf3 between a laptop and an Edge VM on cPouta; the constant baseline uses their medians. For each configuration, task inputs and random seeds remain the same across execution environments and repetitions.

\begin{table}[t]
\centering
\caption{RQ2 configurations. Each configuration runs twice
in each of the four execution environments.}
\label{tab:rq2-configurations}
\small
\setlength{\tabcolsep}{4pt}
\renewcommand{\arraystretch}{1.15}
\begin{tabularx}{\linewidth}{
    @{}>{\raggedright\arraybackslash}p{0.26\linewidth}
    >{\raggedright\arraybackslash}X
    r@{}}
\toprule
Configuration & Network and load & Tasks/run \\
\midrule
Constant baseline
& Constant bandwidth and delay; baseline load
& 5,817 \\

Temporal baseline
& Recorded changes in bandwidth and delay; baseline load
& 5,817 \\

Temporal high load
& The same varying network conditions; higher task load
& 18,282 \\
\bottomrule
\end{tabularx}
\end{table}

\subsubsection{Execution Environments and Acceptance}
Table~\ref{tab:rq2-environments} summarizes the environments used to execute CES. The native run on macOS uses OpenJDK~21 (Corretto), and the three container environments use OpenJDK~17 (Temurin). All runs use a maximum Java heap of 1,536~MiB, Serial garbage collection, and a JVM setting of one active processor. Each run has a 15-minute timeout. The JVM processor setting is separate from the physical CPU allocation. 
\begin{table}[t]
\centering
\caption{RQ2 execution environments and resource allocation.}
\label{tab:rq2-environments}
\small
\setlength{\tabcolsep}{3pt}
\renewcommand{\arraystretch}{1.15}
\begin{tabularx}{\linewidth}{
    @{}>{\raggedright\arraybackslash}p{0.22\linewidth}
    >{\raggedright\arraybackslash}X
    >{\raggedright\arraybackslash}X@{}}
\toprule
Environment & Platform & Resource allocation \\
\midrule
Local
& macOS ARM64
& Managed by the host operating system \\

Docker
& Linux ARM64 container
& Managed by Docker \\

Kubernetes
& CSC Rahti; Linux x86-64
& 1 CPU; 2~GiB memory \\

HPC
& Roihu Slurm/Apptainer; Linux x86-64
& 1 CPU; 4~GiB memory; 30-minute allocation \\
\bottomrule
\end{tabularx}
\end{table}

For each configuration, we compare four CSV files: the overall simulation results, the individual task results, the provider results, and the queue measurements. These files must be byte-identical across all eight runs of that configuration. Acceptance requires all 24 planned runs to be completed and all file comparisons to match exactly, without a numerical tolerance. Host-specific logs, file paths, and execution timestamps are retained separately; execution wall-clock time is also recorded.

\subsection{RQ3: Computational Performance and Scalability}
\label{sec:rq3-methodology}

RQ3 evaluates how total wall-clock time and task throughput scale as the simulated device population grows from 1,000 to 30,000 devices (1000; 5000; 10,000; 20,000; 30,000), comparing CES against both PureEdgeSim (along with its lineage predecessor EISim) and EdgeCloudSim (rev. 83969dd, bundled with CloudSim). Because EdgeCloudSim's bundled CloudSim library requires Java 21 while PureEdgeSim and EISim target Java 17, we run two matched studies on Roihu AMD EPYC 9965 nodes (8 CPUs, 32 GiB memory, 16 GiB maximum Java heap, Apptainer containers, executed sequentially): a Java 17 comparison of CES, EISim and PureEdgeSim (75 runs: 5 seeds $\times$ 5 scales $\times$ 3 simulators) and a paired Java 21 comparison of CES and EdgeCloudSim (50 runs: 5 seeds $\times$ 5 scales $\times$ 2 simulators).

Both studies use the same workload and topology specification: devices generate tasks independently via Poisson processes at 1 task/s per device for 60 simulated seconds (34\% scalar arithmetic, 33\% image processing, 33\% compression) over a star topology of 200 edge servers (one gateway connected directly to 199 peers, all eligible for execution) plus one unused compatibility cloud node, giving 201 infrastructure nodes in total. Each simulator uses the same deterministic round-robin task placement, propagation-only links (1\,ms access hop and 1\,ms server hop each way, with no serialization delay), and runs until all generated tasks finish. In the Java 17 study, each simulator produces an independent deterministic realization of this Poisson model, and post-run gates verify matching total task counts, application shares, per-second arrival distributions, and 100\% task completion. In the paired Java 21 study, both simulators apply a 1-million-instruction minimum per task (matching CloudSim's non-zero cloudlet length constraint), rescale EdgeCloudSim's internal clock to 1\,$\mu$s ticks so integer task lengths preserve sub-millisecond service times, and after each run, a verification check matches every task across both output logs to confirm that all tasks completed with identical inputs, server assignments, and simulated timestamps agreeing within 11\,ns. 

The primary metric is complete process wall-clock time from launch to exit (including container startup, initialization, simulation execution, and output writing), alongside overall throughput (completed tasks per wall-clock second). For each device count $N$, we report medians and min-max ranges across the five seeds, and define CES speedup over baseline $x$ as $\psi_x(N) = \widetilde{T}_x^{\mathrm{wall}}(N) / \widetilde{T}_{\mathrm{CES}}^{\mathrm{wall}}(N)$, computed strictly within the same Java runtime study. In the Java~17 comparison, we also time initialization and topology construction separately to isolate setup costs. In the Java~21 study, we complement wall-clock timing with a separate local Java Flight Recorder profile of EdgeCloudSim (1,000 devices) to identify its execution bottleneck.

\section{Evaluation Results}
\label{sec:evaluation-results}
The following results use the evaluation measurements and the acceptance criteria defined in Section~\ref{sec:evaluation-methodology}.

\subsection{RQ1: Model Fidelity}
\label{sec:results-rq1}

\subsubsection{Server-Timing Experiments}
\begin{figure}[t]
  \centering
  \includegraphics[width=\columnwidth]{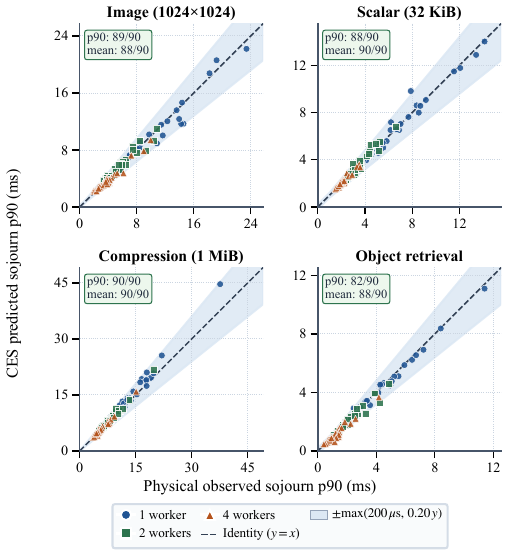}
  \caption{Trace-level observed versus CES-predicted server sojourn p90 (queue arrival to application finish, $f_i - a_i$) across all 90 held-out traces per workload, grouped by worker count ($W \in \{1, 2, 4\}$). Shaded bands show the $\pm\max(200\,\mu\mathrm{s},\,0.20\,y)$ acceptance envelope around the $y = x$ identity line.}
  \label{fig:rq1-server-scatter}
\end{figure}

Table~\ref{tab:rq1-server-comparison} summarizes server-timing fidelity across all four workloads and $W \in \{1, 2, 4\}$ workers at the held-out offered load $\rho = 0.8$. All simulators achieved 100\% within-tolerance hit rates on execution mean and p90, confirming that the calibrated execution profiles accurately describe the held-out runs. Across all four workloads, CES passed 91.1\%-97.8\% of traces on queue mean, 90.0\%-95.6\% on queue p90, 97.8\%-100.0\% on sojourn mean, and 91.1\%-100\% on sojourn p90, maintaining 92.8\%-100\% across every worker group. Figure~\ref{fig:rq1-server-scatter} illustrates trace-level agreement for sojourn time p90 ($S_i = f_i - a_i$): points cluster tightly along the $y = x$ identity line across 1, 2, and 4 workers, with 349 of 360 traces (96.9\%) falling inside the $\pm\max(200\,\mu\mathrm{s}, 0.20\,y)$ tolerance band.

Comparing CES against PureEdgeSim and EdgeCloudSim shows why modeling the full server lifecycle is necessary. Because both baselines run the same non-preemptive $W$-core FIFO scheduler under identical calibrated inputs, they produced identical results. Under standard execution-only calibration, both baselines matched execution times 100\% yet failed all four workloads: they passed only 0.0\%-8.9\% of traces on queue mean, 0.0\% on queue p90, and 33.3\%-52.2\% of checks across worker groups. Releasing a core immediately upon application finish ignores many timing events such as worker post-processing, and collector selection and verification, causing severe underestimation of queuing; their queue p90 mean absolute errors (MAEs) reached $1{,}394.8$-$4{,}570.2\,\mu\mathrm{s}$, which is $7.1\times$ to $15.4\times$ larger than CES ($167.1$-$581.8\,\mu\mathrm{s}$).

Calibrating PureEdgeSim and EdgeCloudSim with full measured worker-slot occupancy ($B_i = u_i - b_i$) substantially reduced these errors, allowing both baselines to pass three of the four workloads (image, scalar, and compression). Even so, CES achieved lower queue p90 MAE on every workload ($407.6$ vs.\ $476.4\,\mu\mathrm{s}$ on image, $172.0$ vs.\ $203.7\,\mu\mathrm{s}$ on scalar, $581.8$ vs.\ $646.5\,\mu\mathrm{s}$ on compression, and $167.1$ vs.\ $253.4\,\mu\mathrm{s}$ on object retrieval) and consistently outperformed them at four workers. On object retrieval, where tasks execute quickly and collector coordination becomes a bottleneck, the occupancy-calibrated baselines failed overall, dropping to 80.0\% on queue p90, 83.3\% and 77.8\% on sojourn mean and p90, and 76.1\% on the four-worker group. In contrast, CES passed every metric (90.0\%-97.8\%) and achieved 92.8\% at four workers. Together, these results show that capturing both post-execution resource retention and serialized collector contention is essential for predicting multi-worker server timing.

\begin{table*}[t]
\centering
\caption{RQ1 server-timing fidelity across four workloads at held-out offered load $\rho = 0.8$ (30 traces per worker count = 90 traces per workload). Metric rows report the percentage of passing traces out of 90; worker-group rows report the percentage of passing checks across all six metrics (30 traces $\times$ 6 = 180 checks). P/EC denotes PureEdgeSim and EdgeCloudSim (identical results). All simulators achieved 100\% on execution mean and p90.}
\label{tab:rq1-server-comparison}
\scriptsize
\setlength{\tabcolsep}{2.2pt}
\begin{tabular}{@{}l@{\hspace{6pt}}rrr@{\hspace{7pt}}rrr@{\hspace{7pt}}rrr@{\hspace{7pt}}rrr@{}}
\toprule
& \multicolumn{3}{c}{\textbf{Image ($1024{\times}1024$)}}
& \multicolumn{3}{c}{\textbf{Scalar (32\,KiB)}}
& \multicolumn{3}{c}{\textbf{Compression (1\,MiB)}}
& \multicolumn{3}{c}{\textbf{Object retrieval (1\,MiB)}} \\
\cmidrule(lr){2-4} \cmidrule(lr){5-7} \cmidrule(lr){8-10} \cmidrule(lr){11-13}
Metric / Worker group
& CES & \shortstack[r]{P/EC\\(occ.)} & \shortstack[r]{P/EC\\(exec.)}
& CES & \shortstack[r]{P/EC\\(occ.)} & \shortstack[r]{P/EC\\(exec.)}
& CES & \shortstack[r]{P/EC\\(occ.)} & \shortstack[r]{P/EC\\(exec.)}
& CES & \shortstack[r]{P/EC\\(occ.)} & \shortstack[r]{P/EC\\(exec.)} \\
\midrule
Queue mean
  & \textbf{91.1\%} & 90.0\% & 0.0\%
  & \textbf{96.7\%} & \textbf{96.7\%} & 8.9\%
  & \textbf{95.6\%} & 92.2\% & 0.0\%
  & \textbf{97.8\%} & 88.9\% & 1.1\% \\
Queue p90 
  & \textbf{94.4\%} & 88.9\% & 0.0\%
  & \textbf{95.6\%} & 88.9\% & 0.0\%
  & \textbf{92.2\%} & 90.0\% & 0.0\%
  & \textbf{90.0\%} & 80.0\% & 0.0\% \\
Sojourn mean 
  & \textbf{97.8\%} & 95.6\% & 2.2\%
  & \textbf{100.0\%} & 98.9\% & 30.0\%
  & \textbf{100.0\%} & 97.8\% & 25.6\%
  & \textbf{97.8\%} & 83.3\% & 0.0\% \\
Sojourn p90
  & \textbf{98.9\%} & 94.4\% & 0.0\%
  & \textbf{97.8\%} & \textbf{97.8\%} & 8.9\%
  & \textbf{100.0\%} & 96.7\% & 1.1\%
  & \textbf{91.1\%} & 77.8\% & 0.0\% \\
Queue p90 MAE ($\mu$s)
  & \textbf{407.6} & 476.4 & 4,570.2
  & \textbf{172.0} & 203.7 & 1,394.8
  & \textbf{581.8} & 646.5 & 4,113.1
  & \textbf{167.1} & 253.4 & 2,575.1 \\
\midrule
1 worker 
  & \textbf{98.3\%} & 97.2\% & 33.3\%
  & \textbf{98.3\%} & \textbf{98.3\%} & 33.3\%
  & \textbf{97.2\%} & 92.2\% & 33.3\%
  & \textbf{100.0\%} & 98.9\% & 33.3\% \\
2 workers 
  & 96.7\% & \textbf{97.2\%} & 33.3\%
  & \textbf{97.2\%} & 96.7\% & 38.3\%
  & 99.4\% & \textbf{100.0\%} & 36.1\%
  & \textbf{95.6\%} & 90.0\% & 33.3\% \\
4 workers 
  & \textbf{96.1\%} & 90.0\% & 34.4\%
  & \textbf{99.4\%} & 96.1\% & 52.2\%
  & \textbf{97.2\%} & 96.1\% & 43.9\%
  & \textbf{92.8\%} & 76.1\% & 33.9\% \\
\bottomrule
\end{tabular}
\par\smallskip
\begin{minipage}{\textwidth}
\footnotesize
P/EC denotes PureEdgeSim and EdgeCloudSim combined (both execute the same non-preemptive $W$-core FIFO discipline and produced identical pass rates and MAEs across all workloads). (occ.) uses measured worker-slot occupancy ($B_i = u_i - b_i$); (exec.) uses application execution duration ($X_i = f_i - s_i$).
\end{minipage}
\end{table*}

\subsubsection{End-to-End Latency}

\begin{figure}[t]
  \centering
  \includegraphics[width=\columnwidth]{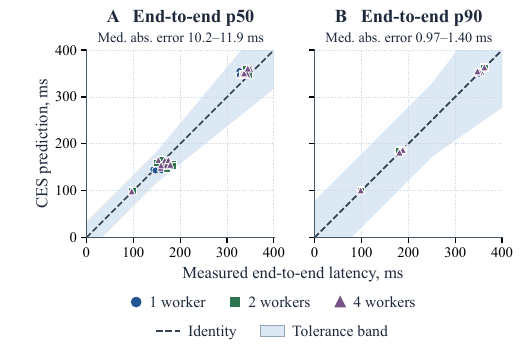}
  \caption{Measured and CES-predicted end-to-end p50 and p90 latency
    for object retrieval across 1, 2, and 4 workers.}
  \label{fig:rq1-network-retrieval}
\end{figure}

Table~\ref{tab:rq1-network-latency} summarizes end-to-end latency fidelity for 1\, MiB object retrieval over the symmetric $100\,\mathrm{Mbit/s}$ cPouta path across 1, 2, and 4 workers. CES with weighted bandwidth sharing passed 100\% of traces (25/25) on all four metrics (end-to-end latency and communication residual at both p50 and p90). Across 1, 2, and 4 workers, median absolute end-to-end errors were $10.20$, $11.93$, and $10.61\,\mathrm{ms}$ at p50 and $0.97$, $1.31$, and $1.40\,\mathrm{ms}$ at p90, with closely matching communication-residual errors ($6.99$-$10.77\,\mathrm{ms}$ at p50 and $1.25$-$1.82\,\mathrm{ms}$ at p90). As shown in Fig.~\ref{fig:rq1-network-retrieval}, all 75 held-out traces fall well inside the acceptance bands and track the identity line closely across isolated, two-request, and four-request bursts.

On four workers, CES also closely reproduced how the 600 held-out tasks were distributed across workers 0-3 (333, 207, 30, and 30 tasks physically vs.\ 330, 210, 30, and 30 in CES); most tasks ran on the first two workers because the collector always assigns the lowest-numbered free worker first, using workers 2 and 3 only when three or four requests overlap simultaneously.

\begin{table}[t]
\centering
\caption{RQ1 end-to-end (E2E) latency and communication-residual (Res.) fidelity for 1\,MiB object retrieval on cPouta across 1, 2, and 4 workers (25 held-out traces and 600 tasks per worker count). Top rows show passing trace rates out of 25; bottom rows show median absolute errors (MedAE, ms) for weighted sharing vs. equal sharing.}
\label{tab:rq1-network-latency}
\scriptsize
\setlength{\tabcolsep}{0pt}
\begin{tabular*}{\columnwidth}{@{\extracolsep{\fill}}lrrrrrr@{}}
\toprule
& \multicolumn{2}{c}{\textbf{1 worker}}
& \multicolumn{2}{c}{\textbf{2 workers}}
& \multicolumn{2}{c}{\textbf{4 workers}} \\
\cmidrule(lr){2-3} \cmidrule(lr){4-5} \cmidrule(l){6-7}
Metric & Weighted & Equal & Weighted & Equal & Weighted & Equal \\
\midrule
E2E p50 (\%)   & \textbf{100\%} & 92\%           & \textbf{100\%} & 96\%           & \textbf{100\%} & \textbf{100\%} \\
E2E p90 (\%)   & \textbf{100\%} & \textbf{100\%} & \textbf{100\%} & \textbf{100\%} & \textbf{100\%} & \textbf{100\%} \\
Res.\ p50 (\%) & \textbf{100\%} & 96\%           & \textbf{100\%} & 96\%           & \textbf{100\%} & \textbf{100\%} \\
Res.\ p90 (\%) & \textbf{100\%} & \textbf{100\%} & \textbf{100\%} & \textbf{100\%} & \textbf{100\%} & \textbf{100\%} \\
\midrule
E2E p50 (ms)   & \textbf{10.20} & 16.56          & 11.93          & \textbf{11.12} & \textbf{10.61} & 11.73 \\
E2E p90 (ms)   & \textbf{0.97}  & 1.12           & 1.31           & \textbf{1.09}  & \textbf{1.40}  & 1.59 \\
Res.\ p50 (ms) & \textbf{6.99}  & 10.76          & \textbf{10.74} & 11.87          & \textbf{10.77} & 11.22 \\
Res.\ p90 (ms) & \textbf{1.44}  & 1.66           & 1.82           & \textbf{1.09}  & \textbf{1.25}  & 1.43 \\
\bottomrule
\end{tabular*}
\end{table}

\begin{figure}[t]
  \centering
  \includegraphics[width=\columnwidth]{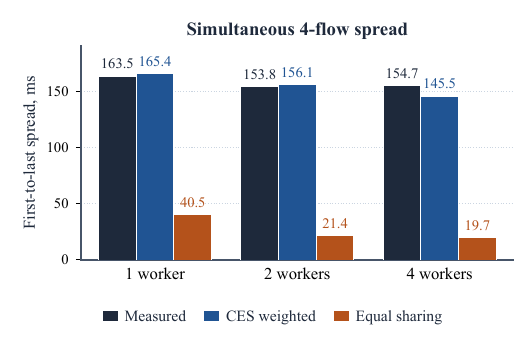}
  \caption{First-to-last completion spread for four simultaneous
    object requests across 1, 2, and 4 workers.}
  \label{fig:rq1-network-retrieval-spread}
\end{figure}

To examine concurrent-flow dynamics directly, Figure~\ref{fig:rq1-network-retrieval-spread} compares the median \emph{completion spread} during simultaneous four-request bursts, defined as the elapsed time from when the earliest response finishes to when the fourth (last) response finishes. On the physical network, overlapping TCP transfers finished staggered rather than together, with measured spreads of $163.5\,\mathrm{ms}$ (1 worker), $153.8\,\mathrm{ms}$ (2 workers), and $154.7\,\mathrm{ms}$ (4 workers). Equal sharing assumed all active transfers progressed at identical rates and predicted near-simultaneous completions only $40.5$, $21.4$, and $19.7\,\mathrm{ms}$ apart, underestimating the physical spread by $123$-$135\,\mathrm{ms}$. In contrast, CES weighted sharing closely reproduced the staggered physical completions across all three worker counts ($165.4$, $156.1$, and $145.5\,\mathrm{ms}$), capturing both trace-level latency accuracy and concurrent-flow sharing dynamics.

\subsection{RQ2: Execution Reproducibility}
\label{sec:results-rq2}
All 24 reproducibility runs were completed successfully. For each configuration, the result files were byte-identical across native local execution, Docker, Rahti (Kubernetes), and Roihu (HPC), covering two architectures (ARM64 and x86-64) and two Java versions.

Table~\ref{tab:rq2-results} summarizes the three configurations. Each configuration was executed twice in each of the four environments, producing eight matching runs. The task count was identical across all runs of each configuration.

\begin{table}[ht]
\centering
\caption{RQ2 reproducibility results. Matching runs have identical result files for the same configuration.}
\label{tab:rq2-results}
\small
\begin{tabular}{@{}lrr@{}}
\toprule
Configuration & Tasks per run & Matching runs \\
\midrule
Constant baseline & 5,817 & 8/8 \\
Temporal baseline & 5,817 & 8/8 \\
Temporal high load & 18,282 & 8/8 \\
\midrule
All configurations & -- & 24/24 \\
\bottomrule
\end{tabular}
\end{table}

\subsection{RQ3: Computational Performance and Scalability}
\label{sec:results-rq3}
CES completes simulations in much less total wall-clock time than PureEdgeSim, EISim, and EdgeCloudSim at every tested device count, while achieving 100\% task completion across all runs, verified workload-distribution equivalence in the Java 17 study, and verified matching individual task results in the paired Java 21 study (Table~\ref{tab:rq3-matched-comparison}).

\begin{table}[H]
\centering
\caption{RQ3 scalability comparison (five-seed medians). Throughput is completed tasks per wall-clock second; speedup is competitor wall time divided by CES wall time within the same Java runtime study.}
\label{tab:rq3-matched-comparison}
\scriptsize
\setlength{\tabcolsep}{0pt}
\begin{tabular*}{\columnwidth}{@{\extracolsep{\fill}}rrrrrr@{}}
\toprule
& \multicolumn{2}{c}{\textbf{Wall time (s)}}
& \multicolumn{2}{c}{\textbf{Throughput (tasks/s)}} & \\
\cmidrule(lr){2-3}\cmidrule(lr){4-5}
\multicolumn{6}{@{}l}{\textbf{CES vs.\ PureEdgeSim (Java~17)}} \\
Devices & CES & PureEdgeSim & CES & PureEdgeSim & Speedup \\
\midrule
1,000  & \textbf{4.64}  & 8.22   & \textbf{13,000} & 7,416 & \textbf{1.77$\times$} \\
5,000  & \textbf{13.30} & 79.68  & \textbf{22,520} & 3,776 & \textbf{5.99$\times$} \\
10,000 & \textbf{16.31} & 94.53  & \textbf{36,762} & 6,350 & \textbf{5.80$\times$} \\
20,000 & \textbf{33.52} & 499.89 & \textbf{35,813} & 2,403 & \textbf{14.91$\times$} \\
30,000 & \textbf{46.16} & 920.00 & \textbf{38,966} & 1,955 & \textbf{19.93$\times$} \\
\midrule[\heavyrulewidth]
\multicolumn{6}{@{}l}{\textbf{CES vs.\ EdgeCloudSim (Java~21)}} \\
Devices & CES & EdgeCloudSim & CES & EdgeCloudSim & Speedup \\
\midrule
1,000  & \textbf{2.78}  & 10.33     & \textbf{21,531} & 5,831 & \textbf{3.71$\times$} \\
5,000  & \textbf{8.90}  & 219.54    & \textbf{33,650} & 1,365 & \textbf{24.66$\times$} \\
10,000 & \textbf{17.24} & 1,791.97  & \textbf{34,772} & 335   & \textbf{103.91$\times$} \\
20,000 & \textbf{35.49} & 8,406.40  & \textbf{33,781} & 143   & \textbf{236.89$\times$} \\
30,000 & \textbf{47.60} & 17,707.76 & \textbf{37,787} & 102   & \textbf{371.99$\times$} \\
\bottomrule
\end{tabular*}
\end{table}

In the Java~17 comparison across all 75 runs (1,000 to 30,000 devices), CES median wall-clock time grew from 4.64\,s to 46.16\,s, compared with 8.22\,s to 920.00\,s for PureEdgeSim (speedup $1.77\times$ to $19.93\times$) and up to 851.87\,s for EISim ($1.70\times$ to $18.45\times$). At 30,000 devices, CES achieved 38,966 tasks/s over total wall-clock time versus 1,955 tasks/s for PureEdgeSim and 2,115 tasks/s for EISim. Separately timed setup stages explain this gap: at 30,000 devices, initialization took 6.64\,s in CES versus 910.77\,s in PureEdgeSim and 843.18\,s in EISim, with topology construction alone consuming 905.35\,s and 837.78\,s compared with 3.37\,s in CES. Both PureEdgeSim and EISim precompute shortest propagation-delay paths across the 201 infrastructure nodes by running one Dijkstra search per ordered pair ($\approx$40,000 searches over the full graph including all device links) and scan the full entity list during registration, whereas CES runs one single-source Dijkstra tree per infrastructure node (201 searches total), producing identical routes without redundant searches. Although EISim and PureEdgeSim spent less time in the isolated event-execution loop (7.67~s and 8.32~s versus 38.55~s in CES at $30,000$ devices), topology setup dominated their total runtime, giving CES a much shorter end-to-end time-to-result at every scale.

\begin{figure}[t]
\centering
\includegraphics[width=\columnwidth]{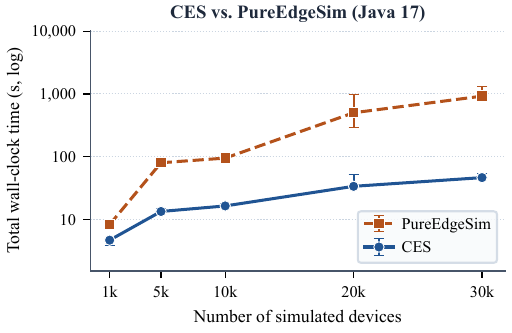}
\caption{Total wall-clock time versus simulated device count for CES and PureEdgeSim (Java~17). Points show medians across five seeds; error bars span minimum to maximum.}
\label{fig:rq3-pureedgesim}
\end{figure}
\begin{figure}[t]
\centering
\includegraphics[width=\columnwidth]{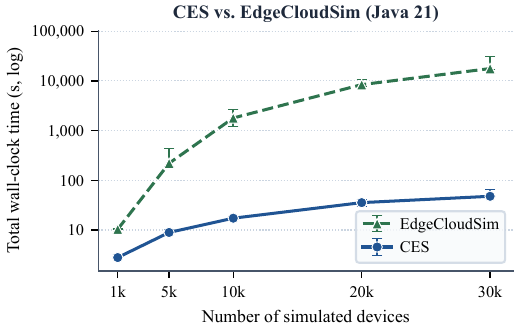}
\caption{Total wall-clock time versus simulated device count for CES and EdgeCloudSim (Java~21). Points show medians across five seeds; error bars span minimum to maximum.}
\label{fig:rq3-edgecloudsim}
\end{figure}

In the paired Java~21 comparison across all 50 runs against EdgeCloudSim (Fig.~\ref{fig:rq3-edgecloudsim}), CES outperformed EdgeCloudSim at every scale, with its advantage growing rapidly with device count: median wall-clock time was 2.78\,s vs.\ 10.33\,s at 1,000 devices ($3.71\times$ speedup; 21,531 vs.\ 5,831 tasks/s), 8.90\,s vs.\ 219.54\,s at 5,000 devices ($24.66\times$ speedup; 33,650 vs.\ 1,365 tasks/s), 17.24\,s vs.\ 1,791.97\,s at 10,000 devices ($103.91\times$ speedup; 34,772 vs.\ 335 tasks/s), 35.49\,s vs.\ 8,406.40\,s at 20,000 devices ($236.89\times$ speedup; 33,781 vs.\ 143 tasks/s), and 47.60\,s vs.\ 17,707.76\,s ($\approx$4.9\,h) at 30,000 devices ($371.99\times$ speedup; 37,787 vs.\ 102 tasks/s). Unlike PureEdgeSim, EdgeCloudSim's bottleneck occurs during simulation execution rather than topology setup: a separate Java Flight Recorder profile of the 1,000-device EdgeCloudSim run showed that 82.9\% of sampled main-thread stacks were spent inside CloudSim's \texttt{bindCloudletToVm} and \texttt{getById} methods. Because EdgeCloudSim's task submission handler (\texttt{submitTaskToVm}) appends tasks to the broker's list and then looks them up sequentially by ID, binding $N$ tasks performs $O(N^2)$ linear ID comparisons, causing EdgeCloudSim's runtime to grow sharply with task count. By avoiding both pairwise setup routing and quadratic task-binding lookups, CES sustains high throughput at scale.

\section{Discussion}
\label{sec:discussion}
\subsection{Implications for Continuum Experimentation}
Our evaluation yields four practical lessons for edge-cloud continuum simulation. First, the RQ1 server-timing comparison against PureEdgeSim and EdgeCloudSim shows that idealized execution-only server models severely underestimate queuing: releasing a worker core immediately upon application finish ignores pre- and post-execution server overheads and inflates queue p90 errors by $7.1\times$-$15.4\times$. Even when baselines are calibrated with full worker-slot occupancy, their accuracy still drops sharply under four-worker object retrieval (76.1\% vs. 92.8\% for CES) because they omit serialized collector contention. Credible multi-worker server simulation therefore requires modeling both full resource retention through result handling and shared control-plane serialization.

Second, the RQ1 end-to-end study demonstrates that equal bandwidth sharing is insufficient for concurrent transfer bursts. Even when equal sharing passes trace-level latency checks using accurately measured link capacity, it assumes overlapping TCP flows finish nearly simultaneously and underestimates four-request completion spread by $123$-$135\,\mathrm{ms}$. Calibrated weighted sharing reproduces both trace-level latency quantiles and staggered concurrent completions without incurring the cost of packet-level simulation.

Third, RQ2 confirms that byte-identical simulation outputs can be maintained across heterogeneous development and production environments (native ARM64 workstations, Docker containers, Kubernetes clusters, and bare-metal x86-64 HPC nodes) when configurations, container images, and artifacts are versioned and verified. This allows researchers to prototype locally and run large parameter sweeps on clusters without environment-induced drift.

Fourth, RQ3 reveals that different continuum simulators hit fundamentally different scaling bottlenecks: PureEdgeSim and EISim are dominated during initialization by redundant pairwise shortest-path searches ($O(V^2)$ over infrastructure nodes), whereas EdgeCloudSim is dominated during simulation execution by $O(N^2)$ linear task-binding lookups in CloudSim's broker. Avoiding both bottlenecks keeps total wall-clock time under one minute at 30,000 devices (1.8\,million tasks), making large-scale sweeps and policy evaluations practical.

\subsection{Limitations}
\paragraph{Calibration cost and transferability}
Calibrating CES for a new target system remains the primary practical cost. In our Roihu study, each workload requires 60 ordinary calibration traces across worker counts plus dedicated component and control probes. On cPouta, end-to-end calibration required link probes, 108 concurrent-burst header captures for sharing weights, and 53-59 server and HTTP traces per configuration. Because execution variability, dispatch overhead, and collector timings depend on CPU architecture, kernel scheduling, runtime version, workload structure, and other factors, calibrated profiles cannot be transferred directly to different machines without recalibration. 
\paragraph{Modeling assumptions and testbed scope}
Our log-additive execution model (Eq.~\eqref{eq:ces-execution-time}) and independent component draws are empirical approximations rather than exhaustive hardware models. In addition, our physical testbeds pinned workers to dedicated cores or vCPUs and used a controlled, symmetric $100\,\mathrm{Mbit/s}$ Ethernet path (TCP CUBIC, HTTP/1.1). Co-located multi-tenant CPU/memory contention, CPU oversubscription, thermal throttling, wireless link jitter, packet loss, cross-traffic interference, and multi-hop network congestion were not evaluated. Finally, our end-to-end validation focused on 1\,MiB object retrieval, and our RQ1 validation targeted timing fidelity rather than physically metered energy, carbon, or monetary accounting.

\subsection{Future Work}
Three directions can substantially extend CES's reach and usability:
\begin{itemize}[leftmargin=*,nosep]
  \item \textbf{Automated and low-cost calibration:} Reducing calibration overhead through adaptive measurement design (collecting probes only until parameter uncertainty meets a target tolerance) and machine-learning surrogate models that predict server and network parameters from hardware descriptors and short pilot runs, enabling faster onboarding and partial transfer across heterogeneous edge nodes.
  \item \textbf{Expanded physical validation:} Extending component calibration and held-out validation to multi-hop topologies, wireless and 5G links with variable bandwidth and packet loss, shared-core multi-tenant contention, and physically metered power, energy, and operational carbon models.
  \item \textbf{Agentic continuum simulation and the agentic economy:} Edge-cloud systems are increasingly moving from static, independent task streams toward autonomous AI agents (including tool-using LLM and physical-AI agents) that generate multi-step, stateful workflows and negotiate dynamically for compute, data, and bandwidth. Building on CES's validated timing engine, deterministic reproducibility, 30,000-device scalability, and pricing/accounting hooks inherited from EISim, future work will support \emph{agentic workloads} (dependent multi-turn reasoning and tool-call graphs) and \emph{agentic economies}, where autonomous client, broker, and provider agents negotiate dynamic tariffs, service-level agreements, and micro-payments across competing continuum tiers. Evaluating market stability, fairness, and incentive compatibility under realistic multi-worker queuing and network contention represents a promising research frontier.
\end{itemize}

\section{Conclusion}
\label{sec:conclusion}
This paper presented CES, a validated, reproducible, and scalable discrete-event simulation platform for the edge-cloud continuum. By combining a calibrated multi-worker server lifecycle model, weighted concurrent network and HTTP sharing, a managed provenance-tracked execution workflow, and a scalable simulation engine, CES addresses core credibility and performance gaps in existing continuum simulators. Across bare-metal HPC and cloud testbeds, CES consistently reproduced held-out server queuing and sojourn latency distributions where idealized baselines failed, produced byte-identical results across four heterogeneous execution environments, and achieved one- to two-order-of-magnitude wall-clock speedups at up to 30,000 devices. With open-source code and archived evaluation artifacts, CES provides a rigorous foundation for repeatable, large-scale edge-cloud continuum studies and emerging agentic continuum systems.

\section*{CRediT authorship contribution statement}
\textbf{Nam Do}: Conceptualization, Methodology, Software, Validation, Formal analysis, Investigation, Data curation, Visualization, Writing -- original draft. \textbf{Rezwan Ahmad Nayreed}: Conceptualization, Software, Investigation, Writing -- review \& editing.  \textbf{Henna Kokkonen}: Software, Investigation, Conceptualization. \textbf{Hassan Mehmood}: Writing -- review \& editing. \textbf{Vu Truong}: Writing -- review \& editing. \textbf{Lauri Lov\'en}: Supervision, Funding acquisition, Conceptualization, Resources, Writing -- review \& editing.

\section*{Data Availability}
The open-source CES software is available on GitHub at \url{https://github.com/Future-Computing-Group/ces-platform}. The archived reproduction package containing the experiment configurations, physical measurements, and verification, analysis, and plotting scripts needed to reproduce the tables and figures in this paper is available on Zenodo at \url{https://doi.org/10.5281/zenodo.24696269}.

\section*{Funding}
This work was supported by the CO2CREATION Strategic Research Council project (grant 372355), by the EC through the HEU NEUROCLIMA project (GA 101137711) and the HEU ARGENTIC project (GA 101298599), the Interreg Aurora ResilientEdge project (Grant Number: 20373282), the ERDF (project numbers A81568, A91867), by Business Finland through the Neural pub/sub research project (diary number 8754/31/2022), and by the Finnish Innovation Fund Sitra through the Physical-AI project (agreement 640102).

\section*{Declaration of Competing Interests}
The authors declare no known competing financial interests or personal relationships that could have appeared to influence the work reported in this paper.

\section*{Acknowledgements}
The authors wish to acknowledge CSC -- IT Center for Science, Finland, for generous computational resources. We thank Rudi Cilibrasi for insightful comments on improving the visualizations and the presentation in Section~\ref{sec:model} in an early draft of this paper. 

\section*{Declaration of generative AI use}
During the preparation of this work, the authors used OpenAI Codex and Gemini 4 Neon to edit the manuscript. The authors reviewed and edited all outputs and take full responsibility for the content of the published article.

\bibliographystyle{elsarticle-num}
\bibliography{ces-refs}

\end{document}